\documentclass[runningheads]{llncs}

\usepackage{pdflscape}
\usepackage[T1]{fontenc}
\usepackage{multirow}
\usepackage{booktabs} 
\usepackage{dirtytalk} 
\usepackage{todonotes} 
\usepackage{graphicx}
\usepackage{tablefootnote} 
\usepackage{amsmath}
\usepackage{amssymb}
\usepackage{siunitx}
\usepackage{bbding}
\begin{document}

\title{
Clustering Malware at Scale: A First Full-Benchmark Study
}

\author{Martin Mocko\inst{1,2} \Envelope \orcidID{0000-0001-8982-0141} \and
Jakub Ševcech\inst{4}\orcidID{0000-0002-6328-1335} \and
Daniela Chudá\inst{2,3}\orcidID{0000-0002-3873-9308}}
\authorrunning{M. Mocko et al.}
%
\institute{Faculty of Information Technology, Brno University of Technology, Brno, Czechia \and
Kempelen Institute of Intelligent Technologies, Bratislava, Slovakia
\and
Faculty of Electrical Engineering and Information Technology, Slovak University of Technology, Bratislava, Slovakia
\and
Swiss Re\\
\email{martin.mocko@kinit.sk}
\email{jakub\_sevcech@swissre.com}
\email{daniela.chuda@stuba.sk}
}

\maketitle              
\begin{abstract}
Recent years have shown that malware attacks still happen with high frequency. Malware experts seek to categorize and classify incoming samples to confirm their trustworthiness or prove their maliciousness. One of the ways in which groups of malware samples can be identified is through malware clustering. Despite the efforts of the community, malware clustering which incorporates benign samples has been under-explored. 
Moreover, despite the availability of larger public benchmark malware datasets, malware clustering studies have avoided fully utilizing these datasets in their experiments, often resorting to small datasets with only a few families.
Additionally, the current state-of-the-art solutions for malware clustering remain unclear.
In our study, we evaluate malware clustering quality and establish the state-of-the-art on Bodmas and Ember - two large public benchmark malware datasets.  Ours is the first study of malware clustering performed on whole malware benchmark datasets. Additionally, we extend the malware clustering task by incorporating benign samples. Our results indicate that incorporating benign samples does not significantly degrade clustering quality.
We find that there are differences in the quality of the created clusters between Ember and Bodmas, as well as a private industry dataset. Contrary to popular opinion, our top clustering performers are K-Means and BIRCH, with DBSCAN and HAC falling behind.



\end{abstract}

\setcounter{footnote}{0} 

\section{Introduction}
\label{sec:intro}
Even in 2025, malware attacks and threats still represent a very relevant and pressing issue. 
Just recently, the security researchers from Human Security\footnote{\url{https://www.humansecurity.com/learn/blog/satori-threat-intelligence-disruption-badbox-2-0/}} have uncovered BADBOX 2.0 - the largest ever uncovered botnet of infected connected TV devices. This botnet contained over 1 million infected devices worldwide and was used for fraud schemes such as click fraud, ad fraud, and unauthorized proxy services. 
The amount of malware (but also of standard benign software) that is created daily makes it impossible to investigate all new incoming samples and assess their maliciousness \emph{manually} by malware experts. Therefore, there is a need to automate the analysis of incoming samples.

When looking at the malware domain from the machine learning perspective, two main ways of approaching the problem become apparent: either utilizing supervised models to classify data into (multiple) classes or utilizing unsupervised models to enable a different outlook on their data based on sample (cluster) similarity. The classification problems can be either binary - malware detection - or into N classes - malware family classification. A similar distinction can be considered for the problems of clustering as well - however, most existing works that perform clustering of malware avoid considering benign samples in their experiments - the assumption is that benign samples have been previously filtered out by another mechanism (for example, a malware detection system).

Our work focuses on the under-researched area of \emph{clustering incoming samples} - under no previous assumptions about their maliciousness. From our perspective, binary program clustering is the correct name for this task. However, we will primarily use the phrase \say{malware clustering} considering that this is only a variant of the task and has a close relationship to the task. Furthermore, some works have used the term malware clustering to mean binary program clustering before this work as well \cite{ali2020scalable,rezaei2021pe}.
Utilizing such an approach makes the final solution more versatile - i.e., there is no assumption or reliance on any filtering approach used to filter out benign samples. This makes the whole solution more flexible in its use since it can, in theory, show interesting relationships between malware and clean software and malware families. The problem is also more complex to solve - the task is not just to differentiate different malware families from each other but also to distinguish these families from the relatively vast space of benign samples.

Utilizing malware clustering on the whole space of all samples (both benign and malicious) thus has the following benefits:
\begin{itemize}
    \item can aid in analysis tasks done by malware experts\footnote{This also applies to regular malware clustering without benign samples.\label{fn:clustadv}},
    \item possibility of generating signatures that cover the whole cluster or multiple clusters\footref{fn:clustadv},
    \item is unbiased by other systems (i.e., does not rely on the effectiveness of other systems for filtering),
    \item provides a different look at sample space via distance/similarity measures, and
    \item has the potential to create compressed datasets for further (ML or other) experimentation 
\end{itemize}



Furthermore, as will be shown in Section \ref{sec:related_work} and to our best knowledge, existing work lacks comparison of malware clustering approaches on large public malware benchmark datasets. 
A few works exist which perform clustering on public benchmark datasets. However, only a subset of the data is usually selected where the number of families is relatively low (i.e., less than 10). At the same time, the whole dataset contains at least two orders of magnitude more malware families. Such approaches do not provide the full picture of the malware clustering quality in scenarios where the number of malware families is much greater. Therefore, a more thorough study on the \emph{full} set of malware benchmark dataset samples is needed.


There are several more issues in the existing works that we, at least partially, try to address with our experiments. First, it is the fact that from the existing literature it is not exactly clear what the state-of-the-art in malware clustering is (explained more in Section \ref{sec:related_work}). 
Second, we identify Homogeneity as the main metric for malware clustering evaluation because when clusters' label cohesion is high, the utility of such clustering is much greater for malware experts.
Besides working with public benchmark malware datasets, we also employ a private industry real-world malware dataset.  Therefore, our findings are also supported by experimental evidence on a real-world dataset. 




The main contribution of this work is multi-fold and can be summarized as follows:
\begin{itemize}
    \item To the best of our knowledge, our work is the first malware clustering study performed on \emph{full} large public malware benchmark datasets (i.e., not on small subsets) - on Bodmas and Ember. Additionally, we also include a real-world private industry dataset in our experiments.
    \item We extend the usual malware clustering task by incorporating benign samples, extending the potential use cases of malware clustering. Our results suggest that the inclusion of benign samples does not have a significant negative effect on the clustering quality. 
    \item Our experiments show that the malware clustering quality is dependent on the (sample) composition of the dataset - with the Ember dataset achieving sub-par results in Homogeneity compared to Bodmas. 
    Our ablation study shows only a small Homogeneity increase when increasing the number of components, ruling out the number of clusters as a significant contributor to the low Ember cluster quality.
    \item Contrary to common assumptions, our experiments show that K-Means and BIRCH outperform other tested clustering approaches for malware clustering, while DBSCAN and Hierarchical Agglomerative Clustering exhibit lower Homogeneity.
\end{itemize}

This paper is organized as follows: Section \ref{sec:related_work} discusses the related work and provides the main ground on which the motivation for our experiments is based. Section \ref{sec:methodology} describes the methodology employed for our experiments. Experiments are conducted and results reported in Section \ref{sec:evaluation}. Discussion about the results can be found in Section \ref{sec:discussion}. We conclude our work in Section \ref{sec:conclusion}.

\section{Related Work}
\label{sec:related_work}

This section summarizes the existing research work relevant to our study. We examine related work in terms of the clustering task being solved, the datasets that are utilized in the study experiments, and from the perspective of clustering algorithms, achieved results and evaluation metrics. We identify gaps in existing research that motivate our work.


\subsection{Clustering Task Perspective}
In this subsection, we will look at existing research from the perspective of the clustering task that the research addresses. We will differentiate between malware clustering where no benign samples were used and refer to it as \say{malware clustering}. If the research task the paper focuses on is clustering of both malware and binary samples, we will refer to it as \say{binary program clustering}. If another task is being solved, it will be pointed out. 
In our literature review, we have found only one work that focuses on the task of binary program clustering \cite{ali2020scalable}. On the other hand, there are a lot more works that focus on malware clustering \cite{faridi2018performance,hu2013mutantx,hu2013duet,perdisci2013scalable,aresu2015clustering,fang2019semi}. Some works do not only perform malware clustering but are also capable of performing new malware family identification/discovery without the need to re-train the clustering approach \cite{rieck2011automatic,pitolli2021malfamaware}. However, they are also in a minority. The usual way new malware family identification is approached is that the solution combines a classification approach that filters out the known malware with a clustering approach that analyzes the rest of the samples.  The latter approach performs the discovery process.

\subsection{Dataset Perspective}




To the best of our knowledge, no existing works perform binary program or malware clustering, \emph{on the full benchmark datasets}. There are some works, however, that use some small subset of benchmark datasets. In the COUGAR paper by Wilkins et al. \cite{wilkins2020cougar}, experiments are performed on a subset of around 5,000 malware (only) samples from the Ember dataset. Their follow-up work extending COUGAR \cite{macaskill2021scaling} utilizes 10,000 malware (only) samples from the Ember dataset. The most recent we found are two very similar works from 2024 by a very similar author collective \cite{jurevckova2024online,jurevckova2024classification}. In these works, the Ember dataset is utilized for experiments. However, it is only a subset that captures seven malware families out of more than 3,000 that can be found in Ember in total (although the number of malware samples is more decent). 

Other identified works that do malware clustering usually utilize a small (i.e., thousands of samples) private industry dataset  \cite{hu2013duet,faridi2018performance} or a dataset from a public malware source that usually is not very large either \cite{rieck2011automatic,perdisci2013scalable}. In most cases, this also means that the number of malware families in the dataset is relatively low (usually between 10 to 100). 
Only a few exceptions can be found. The work of Ali et al. \cite{ali2020scalable} utilizes millions of samples - from 1 million up to 10 million. We are not aware of a similar work. More importantly, the dataset used in the work is not publicly available, and information about the number of malware families is undisclosed. Hu et al. \cite{hu2013mutantx} utilize 132,234 malware samples from a public archive and the two works by Jurečková et al. \cite{jurevckova2024online,jurevckova2024classification} utilize 7 Ember malware families totaling 112,651 malware samples in total. 
Conducting experiments under standardized benchmark conditions, particularly regarding data, helps the community compare proposed approaches and properly establish the state-of-the-art. Thus, the importance of conducting experiments on benchmark datasets cannot be overstated. 

Looking at existing work, it is apparent that there is a wide range of used datasets - public and private, with static or dynamic features, smaller and larger. Most of the existing work utilized static malware datasets. Our work will primarily focus on two public malware benchmark datasets containing the same set of static features.

\subsection{The Perspective of Clustering Algorithms, Results and Evaluation Metrics}
From the perspective of clustering algorithms, many different clustering algorithms have been utilized. From K-Means \cite{macqueen1967some}, DBSCAN \cite{ester1996density}, Hierarchical Agglomerative Clustering (HAC) \cite{maimon2005data}, OPTICS \cite{ankerst1999optics}, BIRCH \cite{zhang1996birch}, clustering based on prototypes, and even LSH Bins. Even when using multiple clustering algorithms in a single research work, the winning algorithm is not always the same. 
Pitolli et al. \cite{pitolli2021malfamaware} proposed using BIRCH to be able to identify new malware families and achieved 81.9\% Adjusted Rand Index. Faridi et al. performed a comparative study with HAC, K-Means++, DBSCAN, Spectral clustering, Affinity Propagation, and LSH Bins and achieved 92.1\% V-Measure. Ali et al. \cite{ali2020scalable} proposed a two-step approach of coarse clustering through K-Means combined with a fine-grained Threshold based Hierarchical Agglomerative Clustering (HAC-T) and achieved 98\% Purity while clustering 66\% of the dataset, therefore regarding the rest 34\% as noise. 
A prototype-based approach was proposed by Hu et al. \cite{hu2013mutantx}, and a Precision of 72\% to 89\% was achieved. 
An evolutionary approach was proposed by Wilkins et al. \cite{wilkins2020cougar} and further extended in a follow-up work by MacAskill et al. \cite{macaskill2021scaling} where Homogeneity of 73\% and 79\% was achieved on a very small Ember subset. 

Another aspect of malware clustering is the number of clusters utilized. The default approach adopted in most works is to utilize the same number of clusters as the reported number of families. However, such an approach highly depends on the quality of the malware family labels. Moreover, multiple works have identified that clustering algorithms sometimes divide malware families into sub-families \cite{hu2013mutantx,hu2013duet}. 

Although many clustering approaches were tested, and even more evaluation metrics were used, it is still unclear what the state-of-the-art malware clustering approach is. One 2021 work \cite{macaskill2021scaling} cited MutantX-S, a work from 2013 \cite{hu2013mutantx}, as state-of-the-art. We have not found any more works mentioning state-of-the-art for malware clustering. Therefore, there is a clear need for a comparison performed on benchmark datasets to bring more clarity to the situation. The existing work makes it non-obvious to know whether the results achieved previously on other, primarily smaller, datasets translate to larger benchmark datasets. 





\section{Malware Clustering on Large Datasets}
\label{sec:methodology}
This section addresses the shortcomings introduced in Sections \ref{sec:intro} and \ref{sec:related_work} and describes the task and methodology employed to perform malware clustering experiments. 
Formally, we define the malware clustering task with the inclusion of benign samples as follows: let $D = \{(x_i, y_i)\}_{i=1}^{n}$ be a dataset of $n$ samples, where $x_i \in \mathbb{R}^{d}$ represents the real-valued feature vector of the $i$-th sample with $d$ dimensions and $y_i \in \mathcal{Y}$ is the ground truth label. The set of all possible labels is defined as $ \mathcal{Y} = \{l_1, l_2, l_3, \hdots ,l_m, b\}$, where $l_1, l_2, l_3, \hdots ,l_m$ correspond to $m$ malware families and $b$ represents the benign class. 
The goal is to partition $\mathbb{R}^{d}$ into a set of clusters $\mathcal{C} = \{C_1, C_2, \hdots\}$, where each $C_j$ is a subset of $\mathbb{R}^{d}$ and $j$ corresponds to the $j$-th cluster.



\subsection{Datasets}
\label{sec:datasets}
Based on the presented related work in Section \ref{sec:related_work}, we defined criteria for a public malware benchmark dataset on which it makes sense to perform the experiments. The criteria are:
\begin{itemize}
    \item large(r) number of samples,
    \item large(r) number of malware families,
    \item availability of malware family labels,
    \item the dataset contains both benign and malware samples,
    \item the dataset contains extracted features (i.e., it is not just a collection of PE samples)
\end{itemize}

For the criteria defined above we have found the following two public benchmark datasets that fit: Bodmas \cite{yang2021bodmas} and Ember \cite{anderson2018ember}. Another large malware benchmark dataset, Sorel \cite{harang2020sorel}, does not fulfill one of the criteria - the availability of malware families. Therefore, we did not include it in this work. We also acquired a real-world industry malware dataset, further on referred to as Security, which the company used to train their actual AI models. It also fulfilled our dataset criteria (besides being publicly available). Therefore we also used it in our experiments. Besides clustering the training sets of the respective datasets, we also predicted clusters for the respective test sets. Ember and Security come with pre-defined test sets, while for Bodmas, a train-test split with a ratio of 80:20 was created.
Table \ref{tab:ds_info} summarizes general information about the respective datasets used for our experiments.

Ember is the first real public malware benchmark dataset, published in 2018 \cite{anderson2018ember}. More specifically, it is a collection of features extracted from various malware and benign samples - malware or benign samples were not made available. Together with the dataset, the authors also published a reusable codebase for extracting static features from PE files, making it a de facto feature standard for publishing malware datasets. Ember can be divided into two distinct datasets. One is an easier (for classification) dataset where the samples were collected up to 2017. The second, where samples were collected up to 2018, is a dataset where the samples were explicitly chosen in a way that the resultant training and test sets would be more challenging for machine learning algorithms to classify, according to the authors\footnote{\url{https://github.com/elastic/ember}}. We are utilizing the more challenging 2018 version of the Ember dataset. Table \ref{tab:ember_features} summarizes features that can be found in the Ember dataset. 

Bodmas \cite{yang2021bodmas}, a dataset published in 2021, tries to improve the public malware benchmark dataset situation in two ways. First, they include timestamps in their dataset. This enables researchers to study concept drift in the malware domain. It also enables researchers to perform their studies on a more fine-grained (i.e., multi-month or multi-year) basis. Second, they provide well-curated malware family information - their family labels were curated by their in-house scripts, and some were labeled manually by domain experts. Bodmas is published with Ember features. The last benefit of Bodmas is that the malware samples are made public - this increases the reusability of their dataset with other sets of features. 

Security is a private malware dataset provided to us by a company from the software security industry. Therefore, we are not allowed to disclose very specific information. The oldest samples date back to 2001, and the latest to 2021. The features are based mostly on static analysis, but some may be based on emulation or dynamic analysis.

\begin{table}[]
\centering
\caption{Description of datasets that met the criteria for experiments conducted in this study. }
\label{tab:ds_info}
\begin{tabular}{lrrrr}
\toprule
\textbf{Dataset}         & \textbf{Year} & \textbf{   \# features} & \textbf{     Dataset size}                                                & \textbf{   \# malware families} \\ \midrule \\[-10pt]
Bodmas          & 2021                  & \num{2381}\tablefootnote{Ember features \label{fn:ember_ftr}} & \num{134435} total                                                  & \num{581}\tablefootnote{determined by malware experts}          \\[3pt]
Ember v2   & 2018                  & \num{2381}\footref{fn:ember_ftr} & \begin{tabular}[c]{@{}c@{}}\num{600000} train\\ \num{200000} test\end{tabular} & $\sim$\num{3000} AVclass                 \\[8pt]
Security & 2021                  & \num{140}        & \begin{tabular}[r]{@{}r@{}}9.6 mil. train \\ \num{100000} test \end{tabular}                                          & $\sim$\num{9800}\tablefootnote{private labeling}      \\ \bottomrule
\end{tabular}

\end{table}

\begin{table}[]
    \centering
    \caption{Description of the \num{2381} Ember features that can be found both in the Ember and Bodmas dataset. Feature position numbers are also provided (with zero-based indexing).}
    \label{tab:ember_features}
    \begin{tabular}{c c}
    \toprule
      \textbf{Feature position}   & \textbf{Feature category} \\
    \midrule 
     0 - 255    & Byte histogram \\
     256 - 511 & Byte entropy histogram \\
     512 - 615 & String extractor \\
     616 - 625 &  General file info \\
     626 - 687    & Header file info \\
     688 - 942    & Section information \\
     943 - 2222    &  Imports information \\
     2223 - 2350    & Exports information \\
     2351 - 2380    & Data directories information \\
     \bottomrule 
    \end{tabular}
\end{table}


\subsection{Clustering Algorithms}
Based on the literature review in Section \ref{sec:related_work}, we have decided to pick several representatives of different types of clustering used in malware clustering research: partitioning-based, density-based, and hierarchy-based. The most popular partitioning-based algorithm in the reviewed malware clustering literature was K-Means \cite{faridi2018performance,ali2020scalable}. Among density-based clustering algorithms, the first place belongs to DBSCAN, which achieved top results in a few works \cite{faridi2018performance,wilkins2020cougar}. Hierarchical Agglomerative Clustering (HAC), our choice for hierarchy-based clustering, also achieved top results \cite{faridi2018performance,ali2020scalable}. Finally, we also chose a hybrid clustering algorithm - BIRCH. It is a hybrid combination of partitioning-based and hierarchy-based clustering. BIRCH was also the clustering algorithm of choice in multiple malware clustering works \cite{pitolli2021malfamaware,perdisci2013scalable,aresu2015clustering}.


\subsection{Representation Learning Techniques}
The Ember feature set has a relatively high dimensionality of \num{2381}. Therefore, a dimensionality reduction technique is appropriate. Existing research also does not specify which dimensionality reduction/representation learning technique is best for malware clustering. Therefore, we chose Principal Component Analysis (PCA), Autoencoder (AE), and UMAP \cite{mcinnes2018umap} for our experiments. UMAP was also used for the COUGAR method \cite{wilkins2020cougar} while PCA and Autoencoder are typical dimensionality reduction techniques used across many machine learning domains. 


\subsection{Experimental Setup}
\begin{figure}[h!] 
  \centering
  \includegraphics[scale=1]{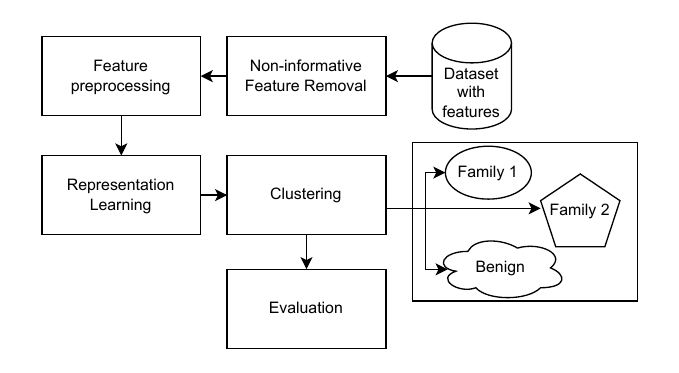}
  \caption{Methodology of our malware clustering experiments}
  \label{fig:methodology}
\end{figure}

As mentioned earlier, we focus on the problem of clustering both malware and benign samples. Figure \ref{fig:methodology} shows a high-level description of our methodology. The datasets identified in Section \ref{sec:datasets} were either already published with a train-test dataset split, or in the case of Bodmas, we created a train-test split with the ratio of 80:20. Our experiment pipeline takes all the samples from the train set of the respective datasets and first performs a feature elimination step. It removes all features that should not be used (i.e., label, family name, and so forth) or  uninformative features and features that harm the representation learning process. During our preliminary phase of experiments, where we transformed features into a lower dimension, we discovered that some Ember features harmed the training process and decided to remove them. This resulted in utilizing \num{2235} features for Ember and Bodmas and \num{135} features in the case of Security. 

The next step of our methodology is data preprocessing. For all three datasets, we perform the same preprocessing steps. We utilize a sequence of three preprocessors from the \texttt{sklearn}\footnote{\url{https://scikit-learn.org/}} library in the following order: RobustScaler (to first remove anomalies), StandardScaler (to standardize our features), and MinMaxScaler (to lower the scale of features into a $[0;1]$ range). 

For representation learning, we used \texttt{sklearn} for PCA, \texttt{tensorflow}\footnote{\url{https://www.tensorflow.org/}} for Autoencoder, and \texttt{umap-learn}\footnote{\url{https://umap-learn.readthedocs.io/}} for training the UMAP representation. The embedding size was chosen to be $10$, providing a great amount of compression as well as making it easier to compute distances compared to the original \num{2381} (\num{2235}) features. 

DBSCAN, HAC, and BIRCH implementations were used from   \texttt{sklearn}. For K-Means, we resorted to utilizing the RAPIDS \texttt{cuML} package to achieve a GPU speedup of the training process. In the end, for some DBSCAN computations, we ended up utilizing the ELKI\footnote{\url{https://elki-project.github.io/}} implementation in Java. The implementation ran much faster and consumed less memory than the \texttt{sklearn} implementation, probably due to the usage of the R-tree index structure. 

We strived to achieve similar conditions for all the clustering algorithms, mainly regarding the number of clusters. Existing work in Section \ref{sec:related_work} shows that clustering algorithms often divide malware families into sub-families. Moreover, the extension of malware clustering by including benign samples poses a need to utilize more clusters than the reported number of malware families, thereby creating space for \say{benign} families. Therefore, we decided to include a wider range of the number of clusters to accommodate the potential need for a higher number of clusters. The number of clusters can be explicitly set for three out of the four algorithms chosen. In this case, we experimented with a wide range of clusters where $n\_clusters = [100, 500, \num{1000},\allowbreak \num{2000},\allowbreak \num{10000},\allowbreak \num{20000}, \allowbreak
\num{30000},\allowbreak \num{50000}]$. These numbers of clusters were utilized for experiments with Ember and Security fully. However, for experiments with Bodmas, we resorted to utilizing only up to \num{2000} clusters. For Bodmas, going too far beyond these numbers is not desirable, as the theorized average number of samples per cluster would be relatively low.

For DBSCAN, the main two parameters affecting the number of created clusters (besides data itself) are the minimum number of points required to form a cluster (referred to as $MinPts$) and the maximum distance between two samples for them to be considered in the neighborhood of each other (referred to as $Epsilon$). It is hard to control DBSCAN's number of created clusters. Therefore, we defined various $Epsilon$ levels which we trained on: $Epsilon = [3, 2, 1, 0.5, 0.3, 0.2, 0.1, 0.05, 0.02]$. Combining three datasets, three representations, four clustering algorithms, and 8-9 cluster sizes resulted in around 300 clustering runs. 

Furthermore, during the initial experimental phase, we found out that HAC tends to use a lot of computer memory due to its large memory complexity. Usually, for a number of clusters greater than \num{2000}, it demanded more than our server's available memory. Therefore, in the case of HAC, we resorted to training on a subset of the training dataset. The rest of the training dataset was then \say{clustered} by training a classification algorithm on the learned clustering assignments of HAC on the train data subset. 

After training the clustering algorithms on the training set, we also perform a prediction for the test set. K-Means and BRICH provide a prediction function in the implementation. For DBSCAN, we implemented the prediction function ourselves based on public consensus\footnote{\url{https://stackoverflow.com/questions/27822752/scikit-learn-predicting-new-points-with-dbscan}}. For HAC, we used the classifier that was used to classify the rest of its training data to predict the cluster assignment on new data. 
Our experiments were run on a CentOS 7 server with 64 CPU cores, 512 GB of RAM, and a 32GB VRAM graphics card.

\subsection{Evaluation Metrics}
The literature review in Section \ref{sec:related_work} showed that different works report different evaluation metrics for their malware clustering results. From our perspective, malware (and benign) clusters are useful when they are largely dominated by the same label (i.e., malware or benign). In the case of malware, they should be dominated by the same malware family. If one malware family was divided into several different sub-families using clustering, this does not constitute a huge problem. There is evidence from existing research that this division into sub-families happens in practice as well \cite{hu2013mutantx,hu2013duet}, thereby creating more clusters than the supposed number of malware families. Therefore, cluster \say{purity} is more important than cluster completeness for our experiments. Consequently, our \emph{primary} metric for experiment evaluation is Homogeneity. It can be defined as: 
\begin{equation}
\label{eq:metrics_homogeneity}
    h = 
    \begin{cases}
        1, & \text{if } H(C,K) = 0 \\
        1 - \frac{H(C|K)}{H(C)}, & \text{otherwise}
    \end{cases}
\end{equation}
Similarly, Completeness can be defined as:
\begin{equation}
\label{eq:metrics_completeness}
    c = 
    \begin{cases}
    1,& \text{if } H(K,C) = 0 \\
    1 - \frac{H(K|C)}{H(K)},              & \text{otherwise}
\end{cases}
\end{equation}

In both equations $H$ corresponds to entropy. The harmonic mean between Homogeneity and Completeness is V-Measure:
\begin{equation}
\label{eq:metrics_vmeasure}
v = 2 \cdot \frac{h \cdot c}{h+c}
\end{equation}

Completeness can inform us about how well the families are localized in one cluster -- the more the families are spread into multiple clusters, the worse the score gets. Additionally, in the extreme case where every sample would become its cluster, we would get a perfect Homogeneity of 100\% and a Completeness of 0\%. However, such a clustering is undesirable and would not be useful for any of the tasks presented in Section \ref{sec:intro} where malware clustering is helpful. Therefore, besides reporting Homogeneity, we also report V-Measure, which we deem a balance between Homogeneity and cluster Completeness. 



\section{Experimental Evaluation}
\label{sec:evaluation}

\subsection{Quantitative Results}
Running around 300 clustering runs yielded a large table of results. We summarize these results using multiple figures and tables. First, we share more general aggregated results about our experiment. Afterwards, results specific to each dataset are presented. If not further specified, we will generally refer to the evaluation metrics achieved on the \emph{train} set of the datasets.

\begin{figure}[h] 
  \centering
  \includegraphics[width=\textwidth]{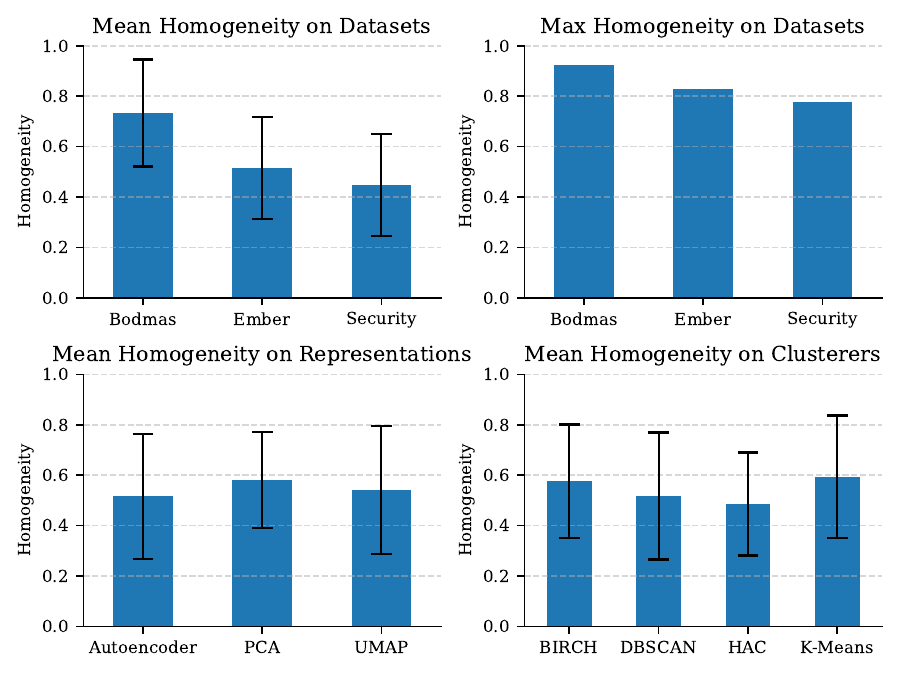} 
  \caption{Aggregated statistics of our experiments (using mean and maximum Homogeneity)}
  \label{fig:aggregated1}
\end{figure}

In the top-left graph of Figure \ref{fig:aggregated1}, we can see that, on average, the clustering algorithms achieved the highest Homogeneity on Bodmas - above 70\%. On Ember, the larger public benchmark malware dataset of the two, the clustering algorithms managed to achieve around 50\% Homogeneity on average. Security, the private real-world industry dataset, achieved a solid third place with Homogeneity being, on average, around 40\%. 

Looking at maximum Homogeneity (in the top-right graph of Figure \ref{fig:aggregated1}) achieved on the datasets, we can see that the difference between mean and maximum Homogeneity can often be quite large, above 20\%. This is related to the relatively large spread of the number of clusters used in our experiments. Table \ref{tab:ds_best} summarizes the best results achieved for the respective datasets. From it, we can see that K-Means consistently beat the other clustering algorithms in the race for best results. The best result overall, and the best result for Bodmas as well, was 92.5\% Homogeneity achieved using UMAP representation. The best achieved Homogeneity on a real-world dataset, Security, was 77.64\% - a relatively good result considering that this is the largest dataset in terms of samples and number of malware families. With regards to Ember, the clustering methods struggled to achieve a clustering quality that would be on par with Bodmas. Nevertheless, the best Homogeneity achieved on Ember was 82.77\%, which is still a very decent result. On all of the datasets, V-Measure stayed relatively low (8\% to 16\%) for the best results - a consequence of penalization through Completeness because of the utilization of more clusters than the reported number of malware families. 

\begin{figure}[h] 
  \centering
  \includegraphics[width=\textwidth]{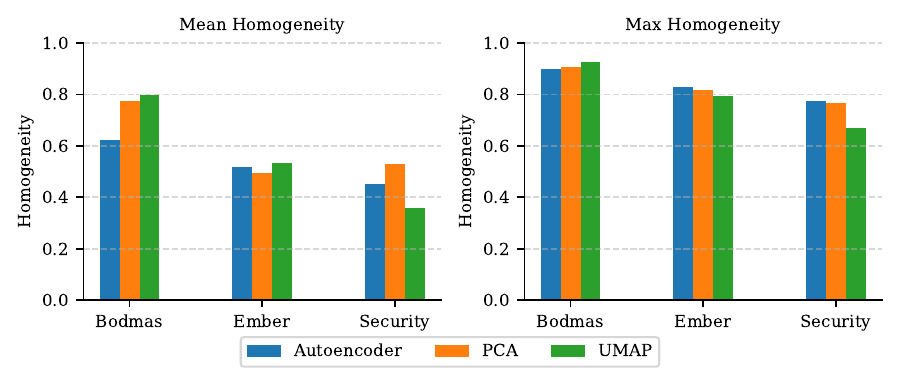}
  \centering
  \caption{Graphs representing mean and maximum Homogeneity on the three utilized representations}
  \label{fig:repre_agg}
\end{figure}

General aggregated results for the three utilized representations (lower-left graph of Figure \ref{fig:aggregated1}) show that PCA, on average, performed better than the other two representations. Interestingly, on average, UMAP and Autoencoder seem to have a very similar performance. A more telling story is when we look at results for the quality of representations grouped by datasets. Figure \ref{fig:repre_agg} shows mean and maximum Homogeneity for the three different representations grouped by datasets. The mean Homogeneity results suggest that UMAP could be considered the best representation for Ember and Bodmas. However, the most important information is which representation could achieve the best results. Indeed, for maximum Homogeneity, we see that PCA and Autoencoder achieved the best results on two out of three datasets. Bodmas shows a different story, with UMAP taking the first place. This could suggest that UMAP provides better representations when the dataset is not too large - as Bodmas is the smallest of the three datasets. 

\begin{figure}[h] 
  \centering
  \includegraphics[scale=1]{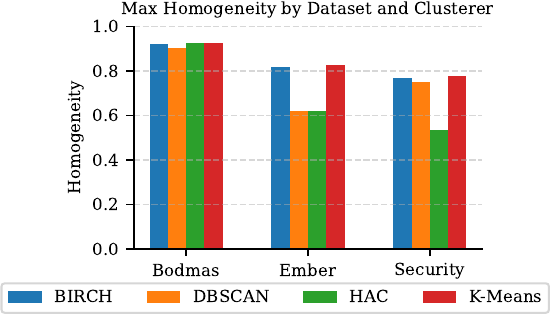}
  \caption{The best results achieved for each dataset and each clustering algorithm}
  \label{fig:clusterer_perf}
\end{figure}

The lower-right graph in Figure \ref{fig:aggregated1} shows the average performance of the four utilized clustering algorithms. We do not show maximum Homogeneity grouped by clustering algorithms because that would only show results achieved on Bodmas (as the results on the other two datasets were worse). The average performance graph shows that K-Means and BIRCH are the two best-performing clustering algorithms in our experiments. DBSCAN is a close third, and HAC was the worst-performing algorithm out of the four. Looking at Figure \ref{fig:clusterer_perf}, we can see that on Bodmas, all the algorithms had very similar high results. For Ember, there is a bigger gap between K-Means and BIRCH as the better performers and HAC and DBSCAN as the less performant algorithms. Clustering the Security dataset resulted in similar maximum performance for three clustering algorithms except for HAC, which performed significantly worse.


\begin{table}[h]
\centering
\caption{The best results, based on the maximum Homogeneity (i.e. best \protect\say{H-train}) achieved, reported for each dataset}
\label{tab:ds_best}
\begin{tabular}{lrrrrrrr}
\toprule
\multicolumn{1}{c}{\textbf{Dataset}} & \multicolumn{1}{c}{\textbf{Represent.}} & \multicolumn{1}{c}{\textbf{Clusterer}} & \multicolumn{1}{c}{\textbf{\# clusters}} & \multicolumn{1}{c}{\textbf{H-train}} & \multicolumn{1}{c}{\textbf{H-test}} & \multicolumn{1}{c}{\textbf{VM-train}} & \multicolumn{1}{c}{\textbf{VM-test}} \\ \midrule
Bodmas & UMAP & K-Means & \num{2000} & 92.50\% & 92.80\% & 15.80\% & 16.01\% \\
Ember & Autoencoder & K-Means & \num{50000} & 82.77\% & 86.76\% & 11.41\% & 13.35\% \\
Security & Autoencoder & K-Means & \num{50000} & 77.64\% & 86.61\% & 10.16\% & 12.02\% \\ \bottomrule
\end{tabular}
\end{table}

\subsection{Ablation Experiment on the Number of Components}
The wide range of results we observed on the datasets inspired us to perform an ablation study. We want to determine what is the influence of the number of components used in the representations on the quality of clustering. Therefore, we performed an ablation experiment in limited conditions. 

We picked K-Means as the clustering algorithm for this experiment - primarily because of its top performance in the main experiment. Regarding representations, UMAP took a long time to train, so we decided to restrict this ablation experiment to only PCA and Autoencoder representations. Besides 10 components that were utilized for the main experiments, we performed experiments with 30 and 50 components. We decided to utilize all three datasets for the experiment because of the large variance in Homogeneity observed between the datasets. Homogeneity was picked as the primary metric to show us the difference in quality (or lack thereof) in this experiment. Due to time constraints, we trained each representation only once. We also ran K-Means only once on the trained representations.

The results of the ablation experiment are reported in Table \ref{tab:ds_ncomp}. For Bodmas, the Homogeneity seemed to increase with every increase in the number of components except for the case of Autoencoder and 30 components. For Ember, Homogeneity (almost) consistently increased with the increase in the number of components, with the exception of \say{H-train-30} where it slightly dropped. The results for the Security dataset are the least consistent among the three datasets. In the case of PCA and 30 components, the Homogeneity decreased a little. On the other hand, for Autoencoder and 30 components, we saw an increase of more than 2\% Homogeneity. For the case of 30 components, the Homogeneity increased in four out of six cases (of combinations of dataset and representation). The Homogeneity increased in all six cases for 50 components (compared to the 10-component baseline).
Conclusions about this experiment will be drawn in Section \ref{sec:discussion}.

\begin{table}[]
\centering
\caption{Results achieved for an ablation experiment where the number of components was increased from 10 to 30 and 50. Homogeneity (\protect\say{H-train-X}) is reported for each dataset and two representations. X stands for the number of components.}
\label{tab:ds_ncomp}
\begin{tabular}{ccrrrr}
\toprule
\textbf{Dataset}          & \textbf{Representation} & \textbf{  H-train-10} & \textbf{  H-train-30} & \textbf{  H-train-50} \\ \midrule
\multirow{2}{*}{Bodmas}   & PCA                     & 90.29\%              & 90.52\%              & 91.03\%              \\
                          & Autoencoder                                    & 89.74\%              & 89.72\%              & 92.02\%              \\ \midrule
\multirow{2}{*}{Ember}    & PCA                     &  81.57\%              & 83.66\%              & 84.70\%              \\
                          & Autoencoder                                        & 82.77\%              & 82.63\%              & 83.64\%              \\ \midrule
\multirow{2}{*}{Security} & PCA                     & 76.69\%              & 76.32\%              & 78.64\%              \\
                          & Autoencoder                                        & 77.64\%              & 79.91\%              & 77.93\% 

                          \\
                          \bottomrule

\end{tabular}
\end{table}


\section{Discussion}
\label{sec:discussion}

\subsection{Interpretation of Results}
The results presented in Section \ref{sec:evaluation} provide multiple insights. 
The results clearly show that Bodmas was the easiest dataset to cluster, achieving above 92\% Homogeneity for the top result. Contrastingly, Ember achieved significantly worse results in terms of Homogeneity (i.e., only reaching around 82\% for the best result on Ember). This constitutes a considerable variance in results. Both of these datasets are public benchmark datasets and share the same set of \num{2381} Ember features. Moreover, they were gathered only roughly 2 years apart - from January 2017 to March 2018 for Ember and from August 2019 to September 2020 for Bodmas \cite{yang2021bodmas}. Therefore, the \emph{main differentiating factor} are the samples inside the datasets. Thus, our key finding is that \emph{malware clustering quality is highly dependent on the composition of the dataset}. 


Our work modified the malware clustering scenario by including benign samples. We can make conclusions based on the results observed in our experiments and in the experiments found in the existing work presented in Section \ref{sec:related_work}.Although often performed on much smaller datasets with much fewer malware families, the existing work showed 98\% Purity, Precision between 72\% to 89\%, 92\% V-Measure, and Homogeneity of 73\% and 79\%. Thus, our observed results for Bodmas, Ember and Security align with the existing literature's observations. We conclude that \emph{the inclusion of benign samples did not have a dramatic effect on the clustering quality in terms of Homogeneity}.

Despite the existing work (Section \ref{sec:related_work}) pointing to DBSCAN and HAC as the better clustering algorithms, we find that K-Means and BIRCH performed better on both the large public malware benchmark datasets and the private real-world industry dataset. In the case of HAC, this can be partially attributed to the necessity of utilizing workarounds to create HAC clustering assignments when using large amounts of clusters. This was needed due to the large memory requirements of the algorithm, which rapidly increase when the number of clusters increases. For DBSCAN, it is not an easy task to find an $Epsilon$ value that will result in a reasonable number of clusters while keeping the level of noise small. Therefore, the search of hyperparameter space would have to be much more extensive in the case of DBSCAN.

In terms of representations, our results show that PCA tends to outperform the other representations on average. However, as was presented in Table \ref{tab:ds_best}, the best results were achieved each on a different representation. The lesson learned is that the utilization of UMAP could bring the best results for smaller malware datasets. In contrast, Autoencoder could help achieve the best results on very large datasets (e.g., Security). 

The ablation experiment that studied how the increase in the number of components influences clustering quality led us to make multiple observations. First, our general observation is that the increasing the number of components helps the clustering quality to some degree. The size of the effect on Homogeneity should be subject to further research. Our results suggest that the increase in Homogeneity could be anywhere between 0.2\% to around 2.5\%. Second, to decrease the uncertainty in the results, we would need to perform the training of the representations multiple times (ideally at least five times). For each trained representation, we would need to perform clustering multiple times. This could increase the time complexity of the experiment by 25 to 100 times. 

Third, one of our primary motivations behind the experiment was also to see the impact of the increase in the number of components on the \emph{Ember dataset}. Even for Ember, for which we have seen worse results in terms Homogeneity than Bodmas (in the main experiment), the increase in the number of components did not help much in increasing Homogeneity. Ideally, we would like to see that the number of components has a greater influence (i.e., at least 5\%, ideally at least 10\%) on the result Homogeneity, but this was not observed. This strengthens our previously stated argument that the influence of the samples (picked into the dataset) on the result Homogeneity cannot be overstated. 

We have established the merit of the usage of a larger amount of clusters than the reported number of malware families in Sections \ref{sec:related_work} and \ref{sec:methodology}. The results presented in Section \ref{sec:evaluation} point to a \emph{potential positive effect of utilizing more clusters than the reported number of malware families}. This can be derived from the substantial increase in Homogeneity when the number of clusters increases. 

However, we acknowledge that this result is nuanced. The Homogeneity metric is equal to 100\% in the extreme case of every sample being in its own cluster. Such clustering would not be useful for any downstream task, but it illustrates the point that Homogeneity should increase with the number of clusters. Therefore, the best way to evaluate this result would be to have a Homogeneity metric adjusted for the \emph{expected increase} when increasing the number of clusters. However, we are not aware of the existence of any such metric. Therefore, there is evidence that it \emph{may} be beneficial to utilize a larger amount of clusters than the reported number of families (when cluster \say{purity} is the main objective).


\subsection{Limitations}
There are multiple limitations of our work. First, our work conducts experiments on datasets whose features are primarily based on static analysis. Static features can suffer when malware is obfuscated using packers and when malware variants are created that are syntactically very different but semantically very similar. 

Both Bodmas and Ember are based solely on static analysis. For Security, a limited number of features are not based solely on static analysis, but most of them are. Therefore, the results presented and conclusions drawn primarily hold for datasets based on static features. We are not able to make conclusions about datasets that are based primarily on dynamic features. 

Additionally, we experiment with two of the three most prominent public malware benchmark datasets. We did not include Sorel \cite{harang2020sorel} primarily due to the unavailability of malware family labels. Moreover, it would be the largest of all datasets, around double the size of Security. In the interest of completeness of the results, performing malware clustering experiments on Sorel would also be beneficial. 

Another issue that can be identified is that Ember and Bodmas were published a few years ago (2018 for Ember and 2021 for Bodmas). Even the Security dataset was created in 2021. Therefore, the clustering results presented here do not necessarily reflect the latest malware landscape, and we acknowledge this. However, we are not aware of any public benchmark up-to-date datasets. This is an ongoing problem in malware research. 

A limitation of this research lies also in the selection of clustering algorithms. While K-Means, DBSCAN, BIRCH, and HAC were chosen due to their prevalence and suitability for malware clustering, other algorithms, such as Canopy clustering or OPTICS, could offer different perspectives on the data.  Furthermore, the hyperparameter tuning (or, more specifically, the exploration of the number of clusters) for each algorithm was conducted within a specific range. A more comprehensive search of the numbers of clusters might reveal further improvements in clustering Homogeneity. Future research could address this limitation by evaluating a broader range of algorithms and employing more sophisticated optimization techniques, mainly in the case of DBSCAN and the search for the ideal $Epsilon$ value.

\subsection{Future Work}
There are multiple possibilities for future extension of our work. Most of them are potential ways how to improve malware clustering quality. For example, our work could be extended by incorporating more clustering approaches. We have already mentioned OPTICS and Canopy clustering in the previous subsection. Approaches such as MutantX-S \cite{hu2013mutantx} and COUGAR \cite{wilkins2020cougar} could also be explored. 

Over last few years, contrastive learning and self-supervised learning have also become increasingly more popular. Moreover, these solutions can often achieve near state-of-the-art performance in the computer vision domain. Therefore, adapting such techniques to help improve malware clustering quality is another possibility of extension. 

Transfer learning is arguably even more popular these days than contrastive learning. Thanks to the shared feature set of the most prominent public malware benchmark datasets, transfer learning is also a viable avenue of exploration. For example, it would be interesting to see how well clustering algorithms trained on Bodmas can predict clusters for Ember or vice-versa. 

A deeper investigation into the problem of worse results on the Ember dataset than on Bodmas is equally important. For instance, merging benign and malware samples from different datasets (e.g., Ember and Bodmas) and repeating the clustering experiments could offer better insights into malware clustering. This approach may help clarify the challenges associated with clustering different parts of these datasets.

Last, our experiments were conducted on features that come from static analysis for the most part. This was also the case in our analyzed existing work in Section \ref{sec:related_work}. Therefore, we see great potential in employing dynamic features that can improve the observed results further.

\section{Conclusion}
\label{sec:conclusion}
Our full-benchmark study with large public (and one private) datasets is the first malware clustering study performed on the \emph{whole} large public malware benchmark datasets, Bodmas and Ember. It does not rely on hand-picked subsets with a limited number of malware families and, therefore, is more realistic in the observed results. Furthermore, we extended the malware clustering task by incorporating benign samples. Our extension can be helpful for malware experts in their daily tasks of analyzing Portable Executables (PEs). It extends the use cases of malware clustering by introducing the ability to analyze new incoming unknown PE samples without a pre-filtering step. Additionally, it creates the potential to create compressed datasets for further (machine learning or other) experimentation. Moreover, we find that the inclusion of benign samples in the malware clustering task does not significantly affect the clustering quality of the solution. 

We have established the state-of-the-art in malware clustering on large public malware benchmark datasets. The best results for Bodmas and Ember were achieved using K-Means and were 92\% and 82\% Homogeneity, respectively. 
Our results also indicate that K-Means and BIRCH achieve higher Homogeneity on these large datasets than DBSCAN and Hierarchical Agglomerative Clustering. Moreover, our findings show that the quality of malware clustering can vary considerably, even if they share the same set of features. This led us to conclude that malware clustering quality is dependent on the composition of the dataset. To examine the results more closely, we performed an ablation study, which increased the number of components of PCA and Autoencoder representations. The ablation results show that the increase in the number of components does not have a substantial effect on clustering Homogeneity, strengthening the result regarding dataset composition. 


Moreover, we also utilized a higher number of clusters than the reported amount of malware families for the dataset. Our results indicate that utilizing a larger number of clusters than the number of families might improve clustering quality - as the results measured using Homogeneity show. However, this takeaway must be treated with caution as Homogeneity is not a metric which is adjusted for the expected increase when increasing the number of clusters. 

Last, we make recommendations regarding the utilized dataset representations - PCA, UMAP, and Autoencoder. Our experiments indicate that UMAP is best suited for smaller malware datasets to achieve the best clustering quality. We recommend utilizing PCA and Autoencoder when working with larger datasets. 

The area of malware clustering still holds potential for future work, mainly in improving the clustering quality. This can be achieved in various ways. The most promising future direction is improving the underlying latent representation of the features. Therefore, our main suggestions for future work are utilizing contrastive learning, self-supervised learning, and transfer learning approaches.


\section*{Acknowledgements and Disclosure of Funding}
An acknowledgement will be added after de-anonymization of the paper.

%
%

\bibliographystyle{splncs04}
\bibliography{biblio}

\end{document}